\documentclass[reprint,amsmath,amssymb,aps, prl]{revtex4-2}

\usepackage{graphicx}
\usepackage{dcolumn}
\usepackage{bm}
\usepackage{xcolor}

\makeatletter
\def\maketitle{
\@author@finish
\title@column\titleblock@produce
\suppressfloats[t]}
\makeatother

\DeclareMathOperator{\Tr}{Tr}

\begin{document}

\title{Surface-induced vortex core restructuring in a spin-triplet superfluid}

\author{Riku Rantanen}
\email{riku.s.rantanen@aalto.fi}
\author{Mikael Huppunen}
\author{Erkki Thuneberg}%
\altaffiliation{QTF Centre of Excellence, Aalto University (AALTO), 00076 Espoo, Finland}
\author{Vladimir Eltsov}
\affiliation{%
 Department of Applied Physics, Aalto University,  PO Box 15100, FI-00076 AALTO, Finland
}%

\date{\today}

\begin{abstract}
Observing the structure of quantized vortices can provide evidence for the pairing nature of a superfluid or superconductor and pinpoint its order parameter. Spin-triplet superfluid $^3$He supports a variety of vortices, calculated and identified so far in bulk fluid. We show numerically that the vortex core in $^3$He is strongly altered near a surface, resulting in a structure inhomogeneous along the vortex line. The effect is asymmetric with respect to the relative orientation of the core order parameter anisotropy axis and the surface normal. Coupling of spin and orbital degrees of freedom in the gradient energy of a triplet order parameter allows  the surface to lift the degeneracy of two bulk vortex states connected by spontaneously broken inversion symmetry of the core.
In a wide range of external conditions, the vortex structure at the surface is found to be completely different from that in bulk. 
As an implication, surface-limited vortex core observations in a triplet-candidate system may not reflect the bulk structure. We propose an experimental verification of the effect by measuring a transition in the vortex structure in thin slabs of superfluid $^3$He-B.
\end{abstract}

\maketitle


{\textit{Introduction}}---Materials that support spin-triplet superconductivity can provide an important platform for hosting topologically protected Majorana modes~\cite{Sato2017,Tanaka2024}, which are beneficial for quantum computing applications. In practice, such materials are rarely found in nature. Recent evidence points toward UTe$_2$ as the leading candidate for intrinsic spin-triplet superconductivity~\cite{Aoki2019,Ran2019,Jiao2020,Hayes2021,Aoki2022,Matsumura2023,Gu2023,Wang2025}, although the exact form of the order parameter is still debated. In addition to measurements of the bulk state of UTe$_2$, the observation of topological defects such as quantized vortices and their detailed structure can be used as an additional constraint to pinpoint the possible pairing states. An unusual asymmetric vortex has been recently observed in scanning tunneling microscope measurements of UTe$_2$~\cite{Sharma2025,Yang2025,Yin2025}, providing further proof of the complex multicomponent nature of the order parameter.

The only system where spin-triplet pairing has been conclusively confirmed and the order parameter definitively established is the neutral $p$-wave superfluid $^3$He. 
The order parameter $\textsf{A}$ is a $3\times 3$ complex matrix, describing the pairing amplitudes on the three spin-triplet states and the three $p$-wave orbital states. 
The multi-component order parameter supports a wide variety of quantized vortices~\cite{Salomaa1987,Thuneberg1987,Parts1995,Blaauwgeers2000,Autti2016,Rantanen2025a}, and their structures are well understood in the bulk. In this Letter we focus on the two types of vortices found in the B phase of $^3$He, the A-phase-core and the double-core vortices, where the former is observed at higher temperatures and pressures and the latter in the rest of the phase diagram~\cite{Pekola1984,Thuneberg1987,Regan2020,Rantanen2024}.

\begin{figure}
    \centering
    \includegraphics[width=\linewidth]{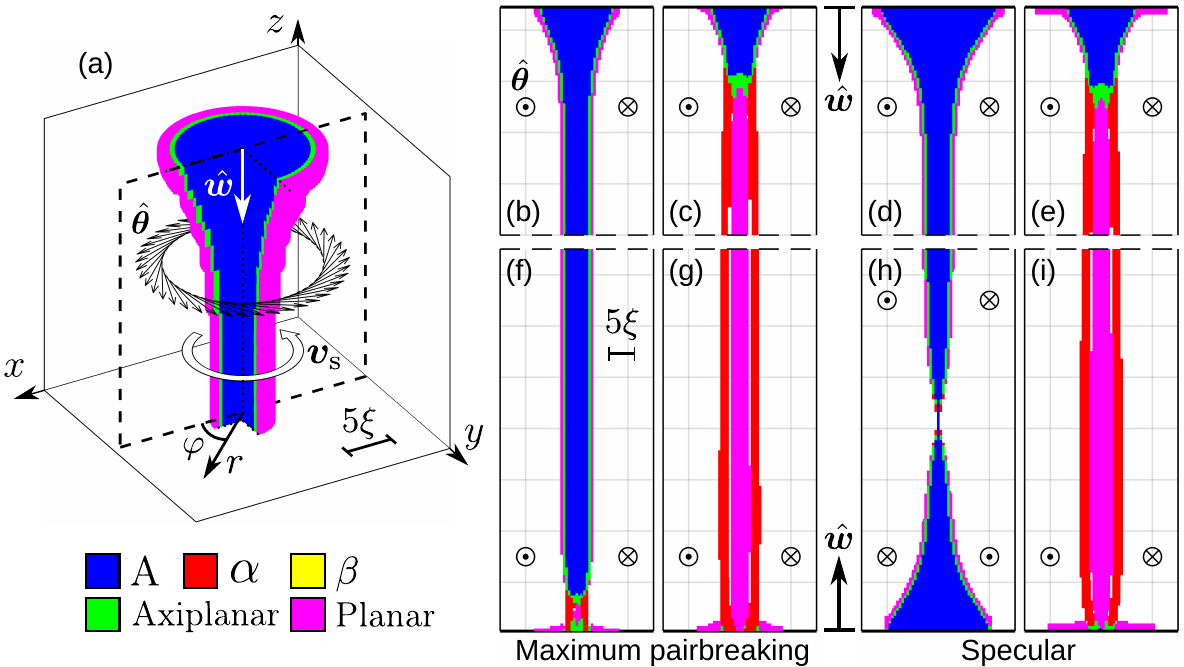}
    \caption{Surface-induced core restructuring. (a) The funnel shape of the A-phase-core vortex in three dimensions. The colors indicate the nearest superfluid phase~\cite{Supplemental,Barton1974}, while the bulk B phase is left transparent. The soft-core orientation $\hat{\bm\theta} = p\hat{\bm\varphi}$  for $p=+1$ and the direction of superfluid velocity $\bm{v}_{\mathrm{s}}$ around the vortex are shown. The dashed line marks the plane where panels (b)-(i) are plotted. (b)-(e) Cross sections along the vortex line in a semi-infinite system with a surface at the top, and (f)-(i) for a surface at the bottom. The orientation of the soft core $\hat{\bm{\theta}}$ is marked with circles with dots (towards viewer) and crosses (away from viewer). The two arrows between panels depict the surface normal $\hat{\bm w}$. In the panels (a-c,f,g) the boundary conditions are maximum pairbreaking, while in panels (d,e,h,i) they are specular. In panels (b,d,f,h) the bulk state is fixed as the A-phase-core vortex and the calculations are performed at $T = 0.85T_{\mathrm{c}}$ and pressure of 29~bar with $\xi=29\,$nm. In panels (c,e,g,i) the bulk state is the double-core vortex, $T = 0.80T_{\mathrm{c}}$ and pressure is 20~bar with $\xi=31\,$nm.}
    \label{fig:semi-infinite}
\end{figure}

A key difference between the experimental methodology for solid state systems like UTe$_2$ and superfluid $^3$He is that in superconductors measurements are typically done using local probes scanning the surface of the material, while in $^3$He observations are often limited to the bulk, for example by measuring the nuclear magnetic resonance response of the whole sample. In this Letter we demonstrate, using numerical simulations of $^3$He, that the vortex core structure observed at the surface can be distinct from the structure in the bulk. It is known that inversion symmetry is spontaneously broken by vortices in $^3$He-B~\cite{Thuneberg1987,Salomaa1987}.
In the bulk this leads to the double degeneracy of the vortex line with given circulation. We have found that at the surface of the sample this degeneracy is lifted and leads to the expansion of the core at one end of the vortex line and contraction at the other (see Fig.~\ref{fig:semi-infinite}). The expanded core is filled with the A phase in a funnel shape independently of the core structure in the bulk. The contracted part usually favors the double-core structure, but a previously unseen symmetric core filled with $\beta$ phase is observed at intermediate temperatures in a thin slab geometry. Remarkably, selection between expanded or contracted core at a particular surface is not related to the direction of the mass current in the vortex, but controlled by the spin anisotropy of the vortex core structure.

The observed vortex restructuring is found to originate from  
coupling
of spin and orbital degrees of freedom in the gradient energy of the order parameter 
(\textit{gradient coupling} for short) revealed by a surface. 
We expect the effect to be general in triplet systems. Our findings are thus crucial for the correct interpretation of vortex measurements in such systems: While making conclusions, one should keep in mind that the vortex structure observed at the surface may be substantially different from that in the bulk.

{\textit{Vortices in bulk}}---In zero magnetic field, two superfluid phases of $^3$He exist in bulk: the A phase at higher pressures and temperatures and the B phase in the rest of the pressure-temperature phase diagram. The order parameter in the B phase is
\begin{equation}
    \textsf{A}= \Delta_{\rm{B}} e^{i\phi}\textsf{R}\,,
    \label{eq:bphase}
\end{equation}
where $\Delta_{\rm{B}}$ is the amplitude and $\textsf{R}$ a rotation matrix that describes the relative orientation of the spin and orbital spaces. It reflects the spontaneously broken symmetry of independent spin and orbital rotations. The rotation matrix is commonly parametrized by a vector $\bm\vartheta$, giving the rotation angle $\vartheta = |\bm{\vartheta}|$ and rotation axis $\hat{\bm{\vartheta}} = \bm{\vartheta}/\vartheta$. Around a quantized vortex in $^3$He-B the phase $\phi$ winds by $2\pi$, which is associated with a superfluid mass flow with velocity $\bm{v}_{\mathrm{s}} = (\hbar/2m)\bm{\nabla}\phi = \nu \hbar/(2mr) \hat{\bm\varphi}$, where 
$m$ is the mass of the $^3$He atom, $(r,\varphi,z)$ are the cylindrical coordinates centered at the vortex, and $\nu = \pm 1$ is the circulation number. At the vortex axis the divergence of the superfluid velocity forces the amplitude $\Delta_{\rm{B}}$ of the B-phase order parameter to zero. The bulk amplitude is recovered at a radius of several Ginzburg-Landau coherence lengths $\xi$. 

In a multicomponent system the cost of condensation energy from the suppression of the bulk order parameter in the vortex core can be compensated by the appearance of other non-rotating components that fill the core and form superfluid phases distinct from bulk. In $^3$He-B there are two such \textit{non-singular} vortex structures which are realized in experiments.

The {\em A-phase-core vortex} is axially symmetric \cite{Salomaa1985}.  Its core is dominated by the A-phase order parameter
\begin{equation}
     \textsf{A}= \Delta_{\rm{A}} \hat{\bm d} (\hat{\bm m}+ i\hat{\bm n})
    \label{eq:aphase}
\end{equation}
where the unit vector $\hat{\bm{d}}$ describes the spin direction of the Cooper pairs and the orthogonal vectors $\hat{\bm{m}}$ and $\hat{\bm{n}}$ define the direction of the orbital angular momentum $\hat{\bm{l}} = \hat{\bm{m}}\times\hat{\bm{n}}$. The {\em double-core vortex} has a more complicated order parameter due to spontaneously broken axial symmetry\cite{Thuneberg1987,Regan2020,Rantanen2025b}. The core is split into two subcores connected by a planar-phase wall. Both of these vortices have spontaneously broken inversion symmetry (reflection in the $xy$ plane) characterized by their parity $p = \pm 1$, which determines the orientation of the spin degrees of freedom in and around the core. The index $p$ is independent of the circulation number $\nu$.
In the bulk, the different values of $p$ are energetically equivalent, but this degeneracy is lifted when the vortex terminates at a surface.

In order to join the superfluid phase in the core to the B phase outside in an energetically optimal way, there is bending of the order parameter on approach to the vortex. This means that $\textsf{R}$ in Eq.~\eqref{eq:bphase} deviates from its bulk value $\textsf{R}(\bm\vartheta_{0})$ in a region known as a soft core~\cite{Hasegawa1985}. The size of the soft core is limited by the spin-orbit interaction and is about $10\,\text{\textmu m}$, much larger than the coherence length $\xi$. In the soft core the rotation matrix 
$\textsf{R}=\textsf{R}(\bm\vartheta_{0})\textsf{R}(\bm \theta)$, where in general $\bm\theta =\theta_\varphi(r,\varphi)\hat{\bm\varphi} + \theta_r(r,\varphi)\hat{\bm r}$. The sign of $\theta_\varphi$ is equal to $p$, which can be used as a practical definition of vortex parity~\cite{Laine2016,Laine2018}. The A-phase-core vortex has a simpler structure with $\theta_r = 0$ and $\theta_\varphi$ depending only on $r$ (Fig.~\ref{fig:semi-infinite}a). 

{\textit{Vortices at the surface}}---The surface provides two important effects for the structure of the vortex core. First, translation symmetry along the vortex in bulk is explicitly broken by the surface, which allows deformations that would be suppressed in a bulk vortex.
Second,
quasiparticle scattering at the surface distorts the order parameter near the boundary~\cite{Ambegaokar1974}. We numerically calculate the structure of a vortex terminating at the surface using Ginzburg-Landau theory of superfluid $^3$He (see Refs.~\cite{Salomaa1985,Thuneberg1987}), with temperature dependent quartic coefficients \cite{wimanthesis,Regan2020,Rantanen2024}. This approach neglects substantial contributions from Fermi-liquid interactions \cite{Silaev15,Nagamura2018}, but we expect the results to be qualitatively correct in a wide temperature range \cite{Supplemental}. The calculation details and boundary conditions \cite{Wiman2015} are given in the Supplemental Material~\cite{Supplemental}.

The calculations reveal a dramatic change in the vortex core structure near the boundary, Fig.~\ref{fig:semi-infinite}. Two types of behavior are found depending on
$p$ and the inwards pointing surface normal $\hat{\bm{w}}$. Because of the connection of $p$ to the orientation of $\bm\theta$ in the soft core, we label a vortex end right-handed if $\hat w_z p> 0$ (as in Fig.~\ref{fig:semi-infinite}f,g,i) and left-handed if $\hat w_z p< 0$ (as in Fig.~\ref{fig:semi-infinite}a-e and Fig.~\ref{fig:semi-infinite}h).

At a left-handed end, the A-phase core expands and forms a "funnel" shape at the boundary. This structure appears independently of the boundary conditions, temperature and pressure in all our simulations. The A-phase-core funnel forms spontaneously even if the bulk vortex state is double-core. At a right-handed end, the A-phase core contracts, while the double core is stable. The contraction may lead to different outcomes depending on conditions, but in many cases it results in a transition to the double-core vortex at the boundary, as shown in the maximum pairbreaking case in Fig.~\ref{fig:semi-infinite}f. 

The extent of the funnel at the left-handed end increases with surface specularity, being the largest for a fully specular surface. Such a surface favors the A-phase funnel so much that at an originally right-handed end we observe a spontaneous reversal of the vortex parity $p$, resulting in a vortex with two left-handed ends (Fig.~\ref{fig:semi-infinite}d,h). The two vortex segments with opposite soft-core orientations are joined via a singular defect (the "o-vortex" \cite{Ohmi1983}), and in the plane of the defect, the soft core disappears completely. The defect is pushed away from the boundary into the bulk until the gradient from the surface funnel becomes negligible. 

\textit{Gradient coupling}---We explain our observations on the basis of the gradient energy for the spin-triplet $p$-wave pairing
\begin{align}
    &f_{\text{grad}} = K_1\nabla_i A_{\mu j}^* \nabla_i A_{\mu j} \nonumber \\ &+ K_2\nabla_i A_{\mu i}^* \nabla_j A_{\mu j} + K_3\nabla_i A_{\mu j}^* \nabla_j A_{\mu i}\,.
    \label{eq:fgrad}
\end{align}
Here the coefficients $K_1$, $K_2$ and $K_3$ are of the same order~\cite{VollhardtWolfle}. The first term with $K_1$ is a generalization of the gradient energy in singlet systems. Triplet-specific physics originates from the $K_2$ and $K_3$ terms. These terms connect spatial coordinates (gradient indices) to the orbital directions (second index of the $\textsf{A}$ matrix)~\cite{Nitta2013}. These terms also couple linear gradients in the orthogonal directions (when $i \ne j$).
A well-known consequence of this
is an induced anisotropy of the order parameter near the surface where, in particular, the A phase with $\hat{\bm{l}}$ along the surface normal becomes energetically favored over the B phase~\cite{Fetter1988,Levitin2013,Saraj2025}. Note that this effect does not explain our observations where the A-phase core shrinks and disappears at the right-handed end of the vortex.

Surface-induced suppression leading to the gradient of the order parameter along the surface normal ($z$ direction) allows the reduction of the energy of the system by forming an appropriate gradient in the direction along the surface ($x$, $y$) owing to the gradient coupling in Eq.~\eqref{eq:fgrad}. A vortex core close to the surface introduces such gradients. The underlying physical picture is that the gradients manifested as the spin currents surrounding the vortex (the soft core) interact with the topological spin currents present at $^3$He-B surfaces~\cite{Zhang1987,Thuneberg2001,Tsutsumi2011,Tsutsumi2012,Wu2013,Mizushima2015,Mizushima2018}
resulting in an overall shift of the energy balance and change of the vortex core size. The direction of spin currents in the vortex depends on the soft core orientation and thus $p$, making the effect asymmetric between left- and right-handed vortex ends. Detailed consideration (see Appendix B in the End Matter and Supplemental Material~\cite{Supplemental}) confirms that this asymmetry is consistent with the numerically calculated structures.

Gradient coupling
appears also on the path between the superfluid phase in the core of a non-singular vortex and the bulk phase. Here the driving gradient is along the radial direction while the energy could be reduced by building a gradient along the vortex. In particular, joining the A-phase core and the bulk B phase requires the $\hat{\bm{l}}$ vector in the A phase to be parallel to the AB interface. As calculations show (see Appendix A in the End Matter), the energy of the vortex soft core can be reduced by tilting the $\hat{\bm{l}}$ vector, and thus the interface, with respect to the vortex axis with the preferred tilt determined by $p$.
This tilt is inhibited by the translation symmetry along the vortex in bulk, as rotating $\hat{\bm{l}}$ would result in a conical shape for the AB interface. Near the surface, however, the core is free to expand (or contract).
In the parameter range relevant to $^3$He-B, this mechanism is found to dominate the observed vortex restructuring. Calculations with no-suppression boundary conditions still reveal formation of the A-phase funnel (see Fig.~\ref{fig:modelcomparison} in End Matter). 

\begin{figure}
    \centering
    \includegraphics[width=\linewidth]{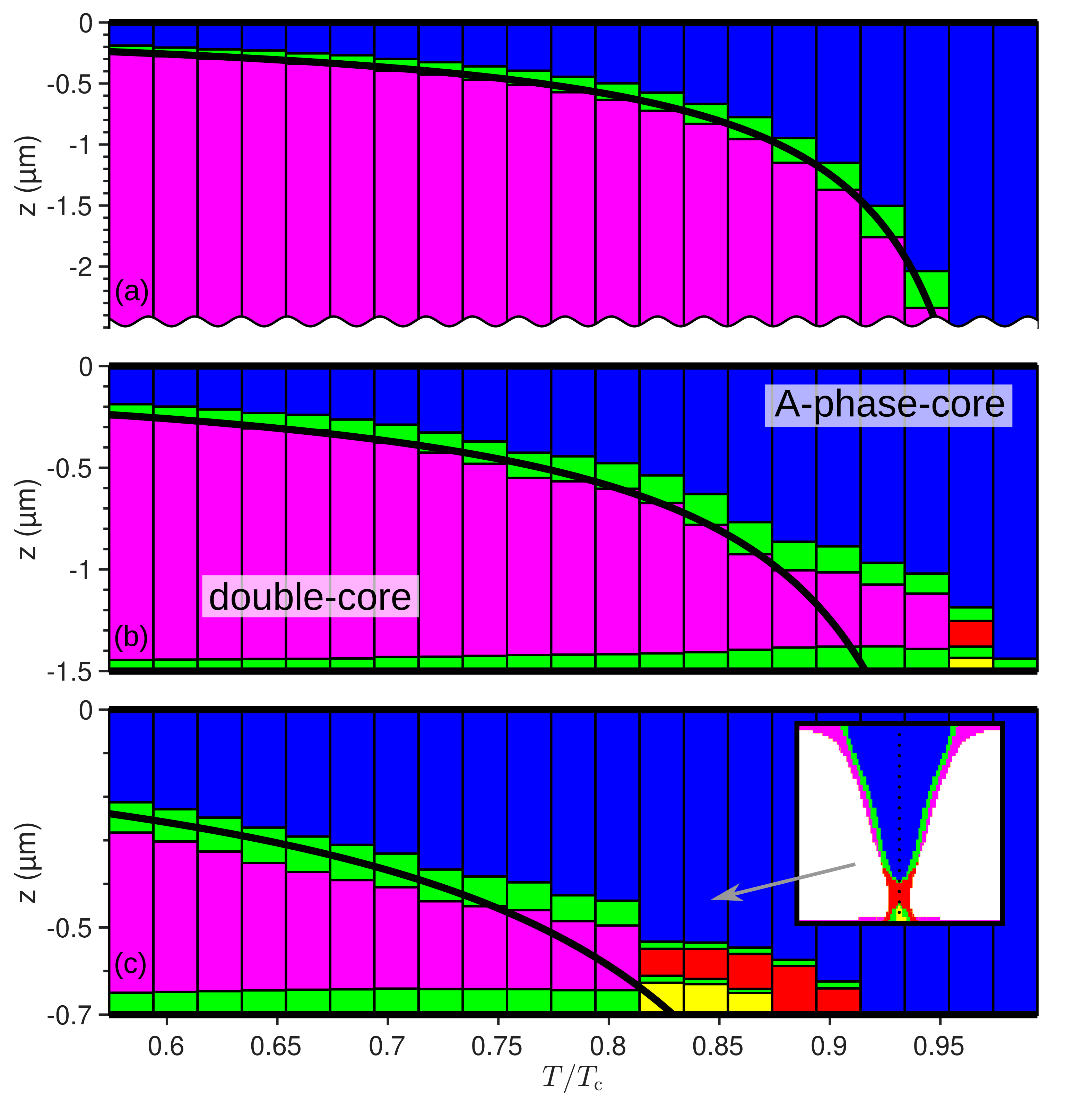}
    \caption{The core structure in a slab as a function of temperature at a pressure of 20~bar. The colored bars show the superfluid phases on the axis of the vortex, with colors corresponding to phases as in Fig.~\ref{fig:semi-infinite}. The panels (a), (b) and (c) correspond to semi-infinite, 1500~nm, and 700~nm slabs, respectively. The thick black line is a fit of $(1-T/T_{\mathrm{c}})^{-1}$ to the height of the A-phase part in the semi-infinite case, with the exact same line overlaid in panels (b) and (c). The inset shows a cross-section of a vortex at $0.84T_{\mathrm{c}}$ in the 700~nm slab, with a squeezed symmetric state at the bottom. Dotted line is the axis along which phase profile is plotted. 
    }
    \label{fig:slab}
\end{figure}

{\textit{Vortices in a slab}}---The effect of surfaces increases
in restricted geometry~\cite{Levitin2019,Heikkinen2020,Shook2020,Heikkinen2024,Heikkinen2025}, when the constriction scale becomes comparable to the size of the funnel. We extend our simulations of a bulk vortex terminating at a boundary to vortices in thin slabs with the experimentally relevant thicknesses~\cite{Levitin2013,Zheng2016,Jiang2023,Shook2024} of 700~nm and 1500~nm. (The bulk-boundary case is called semi-infinite slab below.) The calculations are performed at a pressure of 20~bar and zero field, where the A-phase-core vortex is the energetically favored bulk state only very close to $T_{\mathrm{c}}$~\cite{Regan2020,Rantanen2024}. 

Figure~\ref{fig:slab} shows the nearest superfluid phase along the axis of the vortex, plotted for different temperatures and slab sizes, with maximum pairbreaking boundary conditions. The A phase on the axis signifies the A-phase-core vortex, while planar phase corresponds to the double-core vortex. In the semi-infinite case, Fig.~\ref{fig:slab}(a), the height of the A-phase-core segment increases with temperature as $(1-T/T_{\mathrm{c}})^{-1}$. At low temperatures the vortices in the slabs follow this same behavior, as the size of the slabs is large compared to the height of the A-phase funnel.
At higher temperatures the behavior deviates from the bulk and the A-phase-core segment becomes smaller than in the semi-infinite case. In addition, we observe a distinct core structure at the bottom wall 
at intermediate temperatures, where the double-core structure disappears and instead an axially symmetric state without A phase appears. Close to $T_{\mathrm{c}}$ the A-phase-core funnel extends through the entire thin slab.

\begin{figure}
    \centering
    \includegraphics[width=\linewidth]{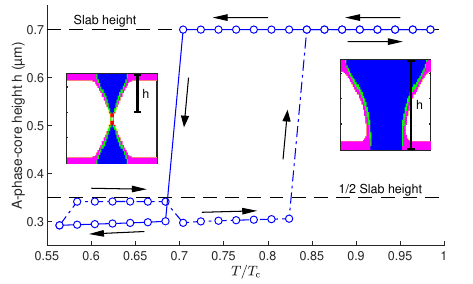}
    \caption{Core transition in a specular 700~nm slab. The data shows the height of the A-phase-core segment on the top boundary as a function of temperature at a pressure of 20~bar. Arrows indicate the direction of the temperature sweep, which shows strong hysteresis. At high temperatures the A-phase-core funnel covers the whole height of the slab (right inset), while at low temperatures the double-funnel structure appears with a defect centered in the middle (left inset).
    }
    \label{fig:specular}
\end{figure}

In the sweeps of the temperature up and down in the maximum pairbreaking slabs we observe only weak hysteresis in the height of the A-phase-core funnel, which we assign to the numerical difficulty of minimizing the energy in our multi-scale domain. A strong hysteresis is observed in slabs with specular boundary conditions, shown in Fig.~\ref{fig:specular}. As in the semi-infinite calculation with specular boundaries in Fig.~\ref{fig:semi-infinite}(d,h), we observe a spontaneous reversal of the soft core at the bottom surface, resulting in a "double-funnel" configuration with a singular defect in the middle of the slab. However, at high temperatures there is not enough room for the double-funnel structure to form and a single defect-free funnel extends through the slab. Sweeping the temperature down from this state results in the A-phase core on the bottom (right-handed) wall contracting, until eventually a defect is formed and shifts to the center of the slab, creating the double-funnel state. The transition between the single- and double-funnel states is hysteretic since it requires formation and removal of the singular defect.

In a thin slab the experimental signature from the double-funnel state is expected to be distinct from the single A-phase-core funnel. For example, relaxation of the NMR signal on vortices occurs via conversion of magnons to light Higgs quasiparticles in the soft vortex cores~\cite{Zavjalov2016}.  In the double-funnel structure, the soft core disappears in the plane of the defect which is expected to produce a signature in the NMR relaxation. The hysteresis in this transition could be used as an additional experimental confirmation of the surface-induced core restructuring. As NMR measurements require the application of magnetic fields, we have calculated the effect of a 500\,G field along the axis of the vortex in a 1500\,nm maximum pairbreaking slab at $T = 0.90T_{\mathrm{c}}$ and a pressure of 20\,bar. The qualitative features of the funnel structure are unchanged, but the height of the A-phase-core segment shrinks by a factor of 0.58. This is expected as the double-core vortex is known to be favored over the A-phase-core vortex in an axial field~\cite{Kasamatsu2019,Rantanen2024}.

{\textit{Discussion}}---While the vortex configurations found in this work, like a double-core vortex crowned with an A-phase-core funnel, may be specific to superfluid $^3$He, the surface-vortex interaction originating from triplet-pairing-relevant terms in the gradient energy is more widely applicable. Unless these terms are suppressed by a symmetry of a particular system, we expect the effects to be general in triplet systems. 

We identify two distinct mechanisms leading to surface-induced core restructuring:
In a spin-triplet system, topological surface spin currents can interact with spin flow around a vortex
and lead to an asymmetric effect
depending on the relative orientation of the surface normal and the spin anisotropy of the vortex. 
In vortices with a topological interface between the core and bulk phases, the core can expand or contract when the translational invariance of the vortex line is
explicitly broken
by the surface. These mechanisms are distict from a trivial modification of the vortex structure following inhomogeneity in the underlying superfluid, e.g.\ near the trap edge in an atomic BEC~\cite{Dalfovo1999}: Our system is uniform with the constant Ginzburg-Landau parameters, while the restructuring effect differentiates between the two ends of a vortex. This effect is also distinct from the spontaneously spatially modulated states suggested in tightly confined $^3$He-B in the slab \cite{Vorontsov2007,Wiman2016,Levitin2019} and cylindrical \cite{Aoyama2014} geometry and from the helical A phase in narrow cylinders~\cite{Wiman2018}, although these effects originate from the same type of gradient coupling. Unlike those states, the core restructuring at the surface is observed without any fine tuning of the confining dimensions, including a single surface limiting bulk volume.

A remarkable outcome of our calculations is that the vortex structure at the surface appears to be independent of the structure in the bulk. For example, on a left-handed vortex end, the A-phase core always appears even at temperatures and pressures where it is unstable in the bulk. If the surface effect is strong enough to reorient the core anisotropy, such as in the double-funnel structure, both ends of the vortex would appear identical. 

While our calculations consider only the spin currents carried by Cooper pairs and detailed consideration of quasiparticle transport at the surface remains a task for the future, it is possible that vortices could be used as a tool to manipulate topological edge states, including spin currents carried by them, via the surface-core interaction.

We propose that the surface-induced core restructuring could be observed in thin slabs of superfluid $^3$He, with the hysteresis in the transition between different configurations dependent on $^4$He surface plating to achieve specular boundary conditions. With specular conditions, our calculations predict a transition between the single- and double-funnel configurations, which is absent in bulk. We believe our findings are also important for the interpretation of the earlier experiments on the core transition in bulk $^3$He-B (see Appendix C in the End Matter).

Our results stress that in solid state systems it is important to probe the bulk state of the vortices in addition to surface measurements. The complete picture could be used to further disambiguate the specific order parameter structure of these systems.

\begin{acknowledgments}
{\textit{Acknowledgments}}---The work was supported by the Research Council of Finland with grant 370532. We acknowledge the computational resources provided by the Aalto Science-IT project.
\end{acknowledgments}

\bibliography{funnelvortex}

@article{Sharma2025,
author = {Sharma, Nileema and Toole, Matthew and McKenzie, James and Cheng, Fangjun and Bordelon, Mitchell M. and Thomas, Sean M. and Rosa, Priscila F. S. and Hsu, Yi-Ting and Liu, Xiaolong},
title = {Observation of Persistent Zero Modes and Superconducting Vortex Doublets in {UTe}$_2$},
journal = {ACS Nano},
volume = {19},
number = {35},
pages = {31539-31550},
year = {2025},
doi = {10.1021/acsnano.5c08406}
}

@article{Yang2025,
    author = {Yang, Zhongzheng and Zheng, Fanbang and Wu, Dingsong and Zhang, Bin-Bin and Li, Ning and Li, Wenhui and Zhang, Chaofan and Zhang, Guang-Ming and Chen, Xi and Chen, Yulin and Yan, Shichao},
    title = {Spectroscopic evidence of symmetry breaking in the superconducting vortices of {UTe}$_2$},
    journal = {National Science Review},
    volume = {12},
    number = {8},
    pages = {nwaf267},
    year = {2025},
    month = {07},
    issn = {2095-5138},
    doi = {10.1093/nsr/nwaf267}
}

@article{Yin2025,
      title = {Yin-{Y}ang vortex on {UTe}$_2$ (011) surface}, 
      author = {Ruotong Yin and Yuanji Li and Zengyi Du and Dengpeng Yuan and Shiyuan Wang and Jiashuo Gong and Mingzhe Li and Ziyuan Chen and Jiakang Zhang and Yuguang Wang and Ziwei Xue and Xinchun Lai and Shiyong Tan and Da Wang and Qiang-Hua Wang and Dong-Lai Feng and Ya-Jun Yan},
      year = {2026},
      journal = {Nat Commun},
      volume = {17},
      pages = {5394},
      doi = {10.1038/s41467-026-72162-9},
      url = {https://www.nature.com/articles/s41467-026-72162-9} 
}

@article{Wang2025,
    author = {Shuqiu Wang and Kuanysh Zhussupbekov and Joseph P. Carroll and Bin Hu and Xiaolong Liu and Emile Pangburn and Adeline Crepieux and Catherine Pepin and Christopher Broyles and Sheng Ran and Nicholas P. Butch and Shanta Saha and Johnpierre Paglione and Cristina Bena and J. C. Séamus Davis and Quangqiang Gu},
    title = {Odd-parity quasiparticle interference in the superconductive surface state of {UTe}$_2$},
    journal = {Nat. Phys.},
    volume = {21},
    pages = {1555},
    year = {2025},
    doi = {10.1038/s41567-025-03000-w}
}

@article{Aoki2019,
author = {Aoki ,Dai and Nakamura ,Ai and Honda ,Fuminori and Li ,DeXin and Homma ,Yoshiya and Shimizu ,Yusei and Sato ,Yoshiki J. and Knebel ,Georg and Brison ,Jean-Pascal and Pourret ,Alexandre and Braithwaite ,Daniel and Lapertot ,Gerard and Niu ,Qun and Vali\v{s}ka ,Michal and Harima ,Hisatomo and Flouquet ,Jacques},
title = {Unconventional Superconductivity in Heavy Fermion {UTe}$_2$},
journal = {Journal of the Physical Society of Japan},
volume = {88},
number = {4},
pages = {043702},
year = {2019},
doi = {10.7566/JPSJ.88.043702}
}

@article{Ran2019,
author = {Sheng Ran  and Chris Eckberg  and Qing-Ping Ding  and Yuji Furukawa  and Tristin Metz  and Shanta R. Saha  and I-Lin Liu  and Mark Zic  and Hyunsoo Kim  and Johnpierre Paglione  and Nicholas P. Butch },
title = {Nearly ferromagnetic spin-triplet superconductivity},
journal = {Science},
volume = {365},
number = {6454},
pages = {684-687},
year = {2019},
doi = {10.1126/science.aav8645}
}

@article{Aoki2022,
author = {Aoki, D. and Brison, J-P and Flouquet, J. and Ishida, K. and Knebel, G. and Tokunaga, Y. and Yanase, Y.},
title = {Unconventional superconductivity in {UTe}$_2$},
journal = {J. Phys.: Condens. Matter},
volume = {34},
pages = {243002},
year = {2022},
doi = {10.1088/1361-648X/ac5863}
}

@article{Hayes2021,
author = {I. M. Hayes  and D. S. Wei  and T. Metz  and J. Zhang  and Y. S. Eo  and S. Ran  and S. R. Saha  and J. Collini  and N. P. Butch  and D. F. Agterberg  and A. Kapitulnik  and J. Paglione },
title = {Multicomponent superconducting order parameter in {UTe}$_2$},
journal = {Science},
volume = {373},
number = {6556},
pages = {797-801},
year = {2021},
doi = {10.1126/science.abb0272}
}

@article{Jiao2020,
    author = {Lin Jiao and Sean Howard and Sheng Ran and Zhenyu Wang and Jorge Olivares Rodriguez and Manfred Sigrist and Ziqiang Wang and Nicholas P. Butch and Vidya Madhavan},
    title = {Chiral superconductivity in heavy-fermion metal {UTe}$_2$},
    journal = {Nature},
    volume = {579},
    pages = {523},
    year = {2020},
    doi = {10.1038/s41586-020-2122-2}
}

@article{Matsumura2023,
author = {Matsumura ,Hiroki and Fujibayashi ,Hiroki and Kinjo ,Katsuki and Kitagawa ,Shunsaku and Ishida ,Kenji and Tokunaga ,Yo and Sakai ,Hironori and Kambe ,Shinsaku and Nakamura ,Ai and Shimizu ,Yusei and Homma ,Yoshiya and Li ,Dexin and Honda ,Fuminori and Aoki ,Dai},
title = {Large Reduction in the $a$-axis {K}night Shift on {UTe}$_2$ with ${T}_{\rm{c}} = 2.1$ {K}},
journal = {Journal of the Physical Society of Japan},
volume = {92},
number = {6},
pages = {063701},
year = {2023},
doi = {10.7566/JPSJ.92.063701}
}

@article{Gu2023,
    author = {Qiangqiang Gu and Joseph P. Carroll and Shuqiu Wang and Sheng Ran and Christopher Broyles and Hasan Siddiquee and Nicholas P. Butch and Shanta R. Saha and Johnpierre Paglione and J. C. Séamus Davis and Xiaolong Liu},
    title = {Detection of a pair density wave state in {UTe}$_2$},
    journal = {Nature},
    volume = {618},
    pages = {921},
    year = {2023},
    doi = {10.1038/s41586-023-05919-7}
}

@article{Sato2017,
doi = {10.1088/1361-6633/aa6ac7},
url = {https://doi.org/10.1088/1361-6633/aa6ac7},
year = {2017},
month = {may},
publisher = {IOP Publishing},
volume = {80},
number = {7},
pages = {076501},
author = {Sato, Masatoshi and Ando, Yoichi},
title = {Topological superconductors: a review},
journal = {Reports on Progress in Physics}
}

@article{Tanaka2024,
    author = {Tanaka, Yukio and Tamura, Shun and Cayao, Jorge},
    title = {Theory of {M}ajorana Zero Modes in Unconventional Superconductors},
    journal = {Progress of Theoretical and Experimental Physics},
    volume = {2024},
    number = {8},
    pages = {08C105},
    year = {2024},
    month = {05},
    issn = {2050-3911},
    doi = {10.1093/ptep/ptae065}
}

@article{Thuneberg1987,
  title = {Ginzburg-{Landau} theory of vortices in superfluid $^{3}${He-B}},
  author = {Thuneberg, E. V.},
  journal = {Phys. Rev. B},
  volume = {36},
  issue = {7},
  pages = {3583--3597},
  numpages = {0},
  year = {1987},
  month = {Sep},
  publisher = {American Physical Society},
  doi = {10.1103/PhysRevB.36.3583},
  url = {https://link.aps.org/doi/10.1103/PhysRevB.36.3583}
}

@article{Thuneberg1991,
  title = {{A-B} interface of superfluid $^{3}\mathrm{He}$ in a magnetic field},
  author = {Thuneberg, E. V.},
  journal = {Phys. Rev. B},
  volume = {44},
  issue = {17},
  pages = {9685--9691},
  numpages = {0},
  year = {1991},
  month = {Nov},
  publisher = {American Physical Society},
  doi = {10.1103/PhysRevB.44.9685},
  url = {https://link.aps.org/doi/10.1103/PhysRevB.44.9685}
}

@article{Thuneberg2001,
   doi       = {10.1023/a:1004898420870},
   title     = {Hydrostatic Theory of Superfluid $^{3}${He-B}},
   author    = {E. V. Thuneberg},
   publisher = {Springer US},
   journal   = {J. Low Temp. Phys.},
   issnp     = {0022-2291},
   issne     = {1573-7357},
   year      = {2001},
   volume    = {122},
   issue     = {5-6},
   pages      = {657--682},

   keys = {texture; dipolar velocity}
}

@article{Regan2020,
  title={Vortex phase diagram of rotating superfluid $^3${He}-{B}},
  author={Regan, R.C. and Wiman, J.J. and Sauls, J.A.},
  journal={Physical Review B},
  volume={101},
  number={2},
  pages={024517},
  year={2020},
  publisher={APS},
  doi = {10.1103/PhysRevB.101.024517},
  url = {https://link.aps.org/doi/10.1103/PhysRevB.101.024517}
}

@article{Rantanen2024,
  title = {Competition of vortex core structures in superfluid $^3${He}-{B}},
  author = {Rantanen, Riku and Eltsov, Vladimir},
  journal = {Phys. Rev. Res.},
  volume = {6},
  issue = {4},
  pages = {043112},
  numpages = {15},
  year = {2024},
  month = {Nov},
  publisher = {American Physical Society},
  doi = {10.1103/PhysRevResearch.6.043112},
  url = {https://link.aps.org/doi/10.1103/PhysRevResearch.6.043112}
}

@article{Salomaa1985,
  title = {Symmetry and structure of quantized vortices in superfluid $^{3}${He-B}},
  author = {Salomaa, M. M. and Volovik, G. E.},
  journal = {Phys. Rev. B},
  volume = {31},
  issue = {1},
  pages = {203--227},
  numpages = {0},
  year = {1985},
  month = {Jan},
  publisher = {American Physical Society},
  doi = {10.1103/PhysRevB.31.203},
  url = {https://link.aps.org/doi/10.1103/PhysRevB.31.203}
}

@article{Schopohl1987,
  title = {Spatial dependence of the order parameter of superfluid $^{3}${He} at the {A-B} phase boundary},
  author = {Schopohl, N.},
  journal = {Phys. Rev. Lett.},
  volume = {58},
  issue = {16},
  pages = {1664--1667},
  numpages = {0},
  year = {1987},
  month = {Apr},
  publisher = {American Physical Society},
  doi = {10.1103/PhysRevLett.58.1664},
  url = {https://link.aps.org/doi/10.1103/PhysRevLett.58.1664}
}

@Inbook{Cross1977,
author="Cross, M. C.",
editor="Trickey, Samuel B.
and Adams, E. Dwight
and Dufty, James W.",
title="Calculation of Surface Energies in {A} and {B} Phases of $^{3}${He}",
bookTitle="Quantum Fluids and Solids",
year="1977",
publisher="Springer US",
address="Boston, MA",
pages="183--194",
isbn="978-1-4684-2418-8",
doi="10.1007/978-1-4684-2418-8_20",
url="https://doi.org/10.1007/978-1-4684-2418-8_20"
}

@article{Hasegawa1985,
    author = {Hasegawa, Yasumasa},
    title = {On Vortex in Superfluid $^{3}${He-B}},
    journal = {Progress of Theoretical Physics},
    volume = {73},
    number = {5},
    pages = {1258-1260},
    year = {1985},
    month = {05},
    issn = {0033-068X},
    doi = {10.1143/PTP.73.1258},
    url = {https://doi.org/10.1143/PTP.73.1258}
}

@article{Zhang1987,
  title = {Order parameter of superfluid $^{3}${He-B} near surfaces},
  author = {Zhang, Weiyi and Kurkij\"arvi, J. and Thuneberg, E. V.},
  journal = {Phys. Rev. B},
  volume = {36},
  issue = {4},
  pages = {1987--1995},
  numpages = {0},
  year = {1987},
  month = {Aug},
  publisher = {American Physical Society},
  doi = {10.1103/PhysRevB.36.1987},
  url = {https://link.aps.org/doi/10.1103/PhysRevB.36.1987}
}

@article{Ohmi1983,
    author = {Ohmi, Tetsuo and Tsuneto, Toshihiko and Fujita, Toshimitsu},
    title = {Core Structure of Vortex Line in $^{3}${He-B}},
    journal = {Progress of Theoretical Physics},
    volume = {70},
    number = {3},
    pages = {647-653},
    year = {1983},
    month = {09},
    issn = {0033-068X},
    doi = {10.1143/PTP.70.647},
    url = {https://doi.org/10.1143/PTP.70.647}
}

@article{Laine2018,
  title = {Spin-wave radiation from vortices in $^3${He}-{B}},
  author = {Laine, S. M. and Thuneberg, E. V.},
  journal = {Phys. Rev. B},
  volume = {98},
  issue = {17},
  pages = {174516},
  numpages = {13},
  year = {2018},
  month = {Nov},
  publisher = {American Physical Society},
  doi = {10.1103/PhysRevB.98.174516},
  url = {https://link.aps.org/doi/10.1103/PhysRevB.98.174516}
}

@article{Silaev15,
  title = {Lifshitz Transition in the Double-Core Vortex in $^3${He}-{B}},
  author = {Silaev, M. A. and Thuneberg, E. V. and Fogelstr\"om, M.},
  journal = {Phys. Rev. Lett.},
  volume = {115},
  issue = {23},
  pages = {235301},
  numpages = {5},
  year = {2015},
  month = {Dec},
  publisher = {American Physical Society},
  doi = {10.1103/PhysRevLett.115.235301},
  url = {https://link.aps.org/doi/10.1103/PhysRevLett.115.235301}
}

@article{Blaauwgeers2000,
    title = {Double-quantum vortex in superfluid $^3${He-A}},
    author = {Blaauwgeers, R. and Eltsov, V. B. and Krusius, M. and Ruohio, J. J. and Schanen, R. and Volovik, G. E.},
    journal = {Nature},
    volume = {404},
    pages = {471-473},
    year = {2000},
    doi = {10.1038/35006583},
    url = {https://www.nature.com/articles/35006583}
}

@article{Rantanen2025a,
    title = {Structure of a single-quantum vortex in $^3${He-A}},
    author = {Rantanen, R. and Thuneberg, E. and Eltsov, V. B.},
    journal = {J. Low Temp. Phys.},
    volume = {220},
    pages = {88-103},
    year = {2025},
    doi = {10.1007/s10909-025-03279-2},
    url = {https://link.springer.com/article/10.1007/s10909-025-03279-2}
}

@article{Rantanen2025b,
    title = {Triple-core structure of the double-core vortex in superfluid $^3${He-B}},
    author = {Rantanen, R.},
    journal = {J. Low Temp. Phys.},
    volume = {222},
    pages = {59},
    year = {2026},
    doi = {10.1007/s10909-026-03394-8},
    url = {https://link.springer.com/article/10.1007/s10909-026-03394-8}
}

@article{Autti2016,
  title = {Observation of Half-Quantum Vortices in Topological Superfluid $^{3}\mathrm{He}$},
  author = {Autti, S. and Dmitriev, V. V. and M\"akinen, J. T. and Soldatov, A. A. and Volovik, G. E. and Yudin, A. N. and Zavjalov, V. V. and Eltsov, V. B.},
  journal = {Phys. Rev. Lett.},
  volume = {117},
  issue = {25},
  pages = {255301},
  numpages = {6},
  year = {2016},
  month = {Dec},
  publisher = {American Physical Society},
  doi = {10.1103/PhysRevLett.117.255301},
  url = {https://link.aps.org/doi/10.1103/PhysRevLett.117.255301}
}

@article{Salomaa1987,
  title = {Quantized vortices in superfluid $^{3}\mathrm{He}$},
  author = {Salomaa, M. M. and Volovik, G. E.},
  journal = {Rev. Mod. Phys.},
  volume = {59},
  issue = {3},
  pages = {533--613},
  numpages = {0},
  year = {1987},
  month = {Jul},
  publisher = {American Physical Society},
  doi = {10.1103/RevModPhys.59.533},
  url = {https://link.aps.org/doi/10.1103/RevModPhys.59.533}
}

@article{Parts1995,
  title = {Phase Diagram of Vortices in Superfluid $^3${He-A}},
  author = {Parts, \"U. and Karim\"aki, J. M. and Koivuniemi, J. H. and Krusius, M. and Ruutu, V. M. H. and Thuneberg, E. V. and Volovik, G. E.},
  journal = {Phys. Rev. Lett.},
  volume = {75},
  issue = {18},
  pages = {3320--3323},
  numpages = {0},
  year = {1995},
  month = {Oct},
  publisher = {American Physical Society},
  doi = {10.1103/PhysRevLett.75.3320},
  url = {https://link.aps.org/doi/10.1103/PhysRevLett.75.3320}
}

@article{Pekola1984,
  title = {Phase Diagram of the First-Order Vortex-Core Transition in Superfluid $^3${He}-{B}},
  author = {Pekola, J. P. and Simola, J. T. and Hakonen, P. J. and Krusius, M. and Lounasmaa, O. V. and Nummila, K. K. and Mamniashvili, G. and Packard, R. E. and Volovik, G. E.},
  journal = {Phys. Rev. Lett.},
  volume = {53},
  issue = {6},
  pages = {584--587},
  numpages = {0},
  year = {1984},
  month = {Aug},
  publisher = {American Physical Society},
  doi = {10.1103/PhysRevLett.53.584},
  url = {https://link.aps.org/doi/10.1103/PhysRevLett.53.584}
}

@article{Laine2016,
    title = {Calculation of {L}eggett-{T}akagi relaxation in vortices of superfluid $^3${He-B}},
    author = {Laine, S. M. and Thuneberg, E. V.},
    journal = {J. Low Temp. Phys.},
    volume = {183},
    pages = {222-229},
    year = {2016},
    doi = {10.1007/s10909-016-1516-x},
    url = {https://link.springer.com/article/10.1007/s10909-016-1516-x}
}

@article{Ambegaokar1974,
  title = {{L}andau-{G}insburg equations for an anisotropic superfluid},
  author = {Ambegaokar, V. and deGennes, P. G. and Rainer, D.},
  journal = {Phys. Rev. A},
  volume = {9},
  issue = {6},
  pages = {2676--2685},
  numpages = {0},
  year = {1974},
  month = {Jun},
  publisher = {American Physical Society},
  doi = {10.1103/PhysRevA.9.2676},
  url = {https://link.aps.org/doi/10.1103/PhysRevA.9.2676}
}

@article{Wiman2015,
  title = {Superfluid phases of $^{3}\mathrm{He}$ in nanoscale channels},
  author = {Wiman, J. J. and Sauls, J. A.},
  journal = {Phys. Rev. B},
  volume = {92},
  issue = {14},
  pages = {144515},
  numpages = {13},
  year = {2015},
  month = {Oct},
  publisher = {American Physical Society},
  doi = {10.1103/PhysRevB.92.144515},
  url = {https://link.aps.org/doi/10.1103/PhysRevB.92.144515}
}

@article{Nitta2013,
  title = {Non-{A}belian quasigapless modes localized on mass vortices in superfluid $^3${He}-{B}},
  author = {Nitta, Muneto and Shifman, Mikhail and Vinci, Walter},
  journal = {Phys. Rev. D},
  volume = {87},
  issue = {8},
  pages = {081702},
  numpages = {7},
  year = {2013},
  month = {Apr},
  publisher = {American Physical Society},
  doi = {10.1103/PhysRevD.87.081702},
  url = {https://link.aps.org/doi/10.1103/PhysRevD.87.081702}
}

@article{Saraj2025,
  title = {Dimensional crossover of superfluid $^{3}\mathrm{He}$ in a magnetic field},
  author = {Saraj, Leyla and Malhotra, Daksh and Muhikira, Aymar and Shook, Alexander J. and Davis, John P. and Boettcher, Igor},
  journal = {Phys. Rev. B},
  volume = {112},
  issue = {17},
  pages = {174509},
  numpages = {34},
  year = {2025},
  month = {Nov},
  publisher = {American Physical Society},
  doi = {10.1103/jfwt-f3bw},
  url = {https://link.aps.org/doi/10.1103/jfwt-f3bw}
}

@article{Levitin2013,
    author = {L. V. Levitin  and R. G. Bennett  and A. Casey  and B. Cowan  and J. Saunders  and D. Drung  and Th. Schurig  and J. M. Parpia },
    title = {Phase Diagram of the Topological Superfluid $^3${He} Confined in a Nanoscale Slab Geometry},
    journal = {Science},
    volume = {340},
    number = {6134},
    pages = {841-844},
    year = {2013},
    doi = {10.1126/science.1233621},
    URL = {https://www.science.org/doi/abs/10.1126/science.1233621}
}

@article{Fetter1988,
    title = {Superfluid density and critical current of $^3${He} in confined geometries},
    author = {Fetter, A. L. and Ullah, S.},
    journal = {J. Low Temp. Phys.},
    volume = {70},
    pages = {515-535},
    year = {1988},
    doi = {10.1007/BF00682163},
    url = {https://link.springer.com/article/10.1007/BF00682163}
}

@article{Wu2013,
  title = {Majorana excitations, spin and mass currents on the surface of topological superfluid $^3${He}-{B}},
  author = {Wu, Hao and Sauls, J. A.},
  journal = {Phys. Rev. B},
  volume = {88},
  issue = {18},
  pages = {184506},
  numpages = {15},
  year = {2013},
  month = {Nov},
  publisher = {American Physical Society},
  doi = {10.1103/PhysRevB.88.184506},
  url = {https://link.aps.org/doi/10.1103/PhysRevB.88.184506}
}

@article{Tsutsumi2011,
  title = {Majorana surface states of superfluid $^{3}\mathrm{He}$ {A} and {B} phases in a slab},
  author = {Tsutsumi, Y. and Ichioka, M. and Machida, K.},
  journal = {Phys. Rev. B},
  volume = {83},
  issue = {9},
  pages = {094510},
  numpages = {12},
  year = {2011},
  month = {Mar},
  publisher = {American Physical Society},
  doi = {10.1103/PhysRevB.83.094510},
  url = {https://link.aps.org/doi/10.1103/PhysRevB.83.094510}
}

@article{Tsutsumi2012,
author = {Tsutsumi ,Yasumasa and Machida ,Kazushige},
title = {Edge Current due to {M}ajorana Fermions in Superfluid $^3${He} {A}- and {B}-Phases},
journal = {Journal of the Physical Society of Japan},
volume = {81},
number = {7},
pages = {074607},
year = {2012},
doi = {10.1143/JPSJ.81.074607},
URL = {https://doi.org/10.1143/JPSJ.81.074607}
}

@article{Mizushima2015,
doi = {10.1088/0953-8984/27/11/113203},
url = {https://doi.org/10.1088/0953-8984/27/11/113203},
year = {2015},
month = {mar},
publisher = {IOP Publishing},
volume = {27},
number = {11},
pages = {113203},
author = {Mizushima, Takeshi and Tsutsumi, Yasumasa and Sato, Masatoshi and Machida, Kazushige},
title = {Symmetry protected topological superfluid $^3${He-B}},
journal = {Journal of Physics: Condensed Matter}
}

@article{Mizushima2018,
    author = {Mizushima, T. and Machida, K.},
    title = {Multifaceted properties of {A}ndreev bound states: interplay of symmetry and topology},
    journal = {Philosophical Transactions of the Royal Society A: Mathematical, Physical and Engineering Sciences},
    volume = {376},
    number = {2125},
    pages = {20150355},
    year = {2018},
    month = {06},
    issn = {1364-503X},
    doi = {10.1098/rsta.2015.0355},
    url = {https://doi.org/10.1098/rsta.2015.0355}
}

@article{Jiang2023,
  title = {Superfluid $^3${He}-{B} surface states in a confined geometry probed by a microelectromechanical oscillator},
  author = {Jiang, W. G. and Barquist, C. S. and Gunther, K. and Lee, Y. and Chan, H. B.},
  journal = {Phys. Rev. B},
  volume = {108},
  issue = {2},
  pages = {024508},
  numpages = {9},
  year = {2023},
  month = {Jul},
  publisher = {American Physical Society},
  doi = {10.1103/PhysRevB.108.024508},
  url = {https://link.aps.org/doi/10.1103/PhysRevB.108.024508}
}

@article{Heikkinen2020,
    title = {Fragility of surface states in topological superfluid $^3${He}},
    author = {Heikkinen, P. J. and Casey, A. and Levitin, L. V. and Rojas, X. and Vorontsov, A. and Sharma, P. and Zhelev, N. and Parpia, J. M. and Saunders, J.},
    journal = {Nat Commun},
    volume = {12},
    pages = {1574},
    year = {2021},
    doi = {10.1038/s41467-021-21831-y},
    url = {https://www.nature.com/articles/s41467-021-21831-y}
}

@article{Shook2020,
  title = {Stabilized Pair Density Wave via Nanoscale Confinement of Superfluid $^{3}\mathrm{He}$},
  author = {Shook, A. J. and Vadakkumbatt, V. and Senarath Yapa, P. and Doolin, C. and Boyack, R. and Kim, P. H. and Popowich, G. G. and Souris, F. and Christani, H. and Maciejko, J. and Davis, J. P.},
  journal = {Phys. Rev. Lett.},
  volume = {124},
  issue = {1},
  pages = {015301},
  numpages = {6},
  year = {2020},
  month = {Jan},
  publisher = {American Physical Society},
  doi = {10.1103/PhysRevLett.124.015301},
  url = {https://link.aps.org/doi/10.1103/PhysRevLett.124.015301}
}

@article{Zheng2016,
  title = {Anomalous Damping of a Microelectromechanical Oscillator in Superfluid $^{3}\mathrm{He}$-{B}},
  author = {Zheng, P. and Jiang, W. G. and Barquist, C. S. and Lee, Y. and Chan, H. B.},
  journal = {Phys. Rev. Lett.},
  volume = {117},
  issue = {19},
  pages = {195301},
  numpages = {5},
  year = {2016},
  month = {Nov},
  publisher = {American Physical Society},
  doi = {10.1103/PhysRevLett.117.195301},
  url = {https://link.aps.org/doi/10.1103/PhysRevLett.117.195301}
}

@article{Levitin2019,
  title = {Evidence for a Spatially Modulated Superfluid Phase of $^{3}\mathrm{He}$ under Confinement},
  author = {Levitin, Lev V. and Yager, Ben and Sumner, Laura and Cowan, Brian and Casey, Andrew J. and Saunders, John and Zhelev, Nikolay and Bennett, Robert G. and Parpia, Jeevak M.},
  journal = {Phys. Rev. Lett.},
  volume = {122},
  issue = {8},
  pages = {085301},
  numpages = {6},
  year = {2019},
  month = {Feb},
  publisher = {American Physical Society},
  doi = {10.1103/PhysRevLett.122.085301},
  url = {https://link.aps.org/doi/10.1103/PhysRevLett.122.085301}
}

@article{Shook2024,
  title = {Surface State Dissipation in Confined $^3${He}-{A}},
  author = {Shook, Alexander J. and Varga, Emil and Boettcher, Igor and Davis, John P.},
  journal = {Phys. Rev. Lett.},
  volume = {132},
  issue = {15},
  pages = {156001},
  numpages = {7},
  year = {2024},
  month = {Apr},
  publisher = {American Physical Society},
  doi = {10.1103/PhysRevLett.132.156001},
  url = {https://link.aps.org/doi/10.1103/PhysRevLett.132.156001}
}

@article{Heikkinen2024,
    title = {Nanofluidic platform for studying the first-order phase transitions in superfluid helium-3},
    author = {Heikkinen, P. J. and Eng, N. and Levitin, L. V. and Rojas, X. and Singh, A. and Autti, S. and Haley, R. P. and Hindmarsh, M. and Zmeev, D. E. and Parpia, J. M. and Casey, A. and Saunders, J.},
    journal = {J. Low Temp. Phys.},
    volume = {215},
    pages = {477-494},
    year = {2024},
    doi = {10.1007/s10909-024-03146-6},
    url = {https://link.springer.com/article/10.1007/s10909-024-03146-6}
}

@article{Heikkinen2025,
  title = {Chiral Superfluid Helium-3 in the Quasi-Two-Dimensional Limit},
  author = {Heikkinen, Petri J. and Levitin, Lev V. and Rojas, Xavier and Singh, Angadjit and Eng, Nathan and Casey, Andrew and Saunders, John and Vorontsov, Anton and Zhelev, Nikolay and Sebastian, Abhilash Thanniyil and Parpia, Jeevak M.},
  journal = {Phys. Rev. Lett.},
  volume = {134},
  issue = {13},
  pages = {136001},
  numpages = {8},
  year = {2025},
  month = {Mar},
  publisher = {American Physical Society},
  doi = {10.1103/PhysRevLett.134.136001},
  url = {https://link.aps.org/doi/10.1103/PhysRevLett.134.136001}
}

@article{Zavjalov2016,
    title = {Light {H}iggs channel of the resonant decay of magnon condensate in superfluid $^3${He-B}},
    author = {Zavjalov, V. V. and Autti, S. and Eltsov, V. B. and Heikkinen, P. J. and Volovik, G. E.},
    journal = {Nat Commun},
    volume = {7},
    pages = {10294},
    year = {2016},
    doi = {10.1038/ncomms10294},
    url = {https://www.nature.com/articles/ncomms10294}
}

@article{Kasamatsu2019,
  title = {Effects of a magnetic field on vortex states in superfluid $^3${He}-{B}},
  author = {Kasamatsu, Kenichi and Mizuno, Ryota and Ohmi, Tetsuo and Nakahara, Mikio},
  journal = {Phys. Rev. B},
  volume = {99},
  issue = {10},
  pages = {104513},
  numpages = {11},
  year = {2019},
  month = {Mar},
  publisher = {American Physical Society},
  doi = {10.1103/PhysRevB.99.104513},
  url = {https://link.aps.org/doi/10.1103/PhysRevB.99.104513}
}

@BOOK{VollhardtWolfle,
   author       = {D. Vollhardt and P. Wölfle},
   year         = 1990,
   title        = {The Superfluid Phases of Helium 3},
   publisher    = {Taylor\&Francis},
   address = "London",
   doi = {10.1201/b12808}
}

@article{Kondo1991,
  title = {Direct observation of the nonaxisymmetric vortex in superfluid $^3${He}-{B}},
  author = {Kondo, Y. and Korhonen, J. S. and Krusius, M. and Dmitriev, V. V. and Mukharsky, Y. M. and Sonin, E. B. and Volovik, G. E.},
  journal = {Phys. Rev. Lett.},
  volume = {67},
  issue = {1},
  pages = {81--84},
  numpages = {0},
  year = {1991},
  month = {Jul},
  publisher = {American Physical Society},
  doi = {10.1103/PhysRevLett.67.81},
  url = {https://link.aps.org/doi/10.1103/PhysRevLett.67.81}
}

@article{Hakonen1983,
  title = {Magnetic Vortices in Rotating Superfluid $^3${He}-{B}},
  author = {Hakonen, P. J. and Krusius, M. and Salomaa, M. M. and Simola, J. T. and Bunkov, Yu. M. and Mineev, V. P. and Volovik, G. E.},
  journal = {Phys. Rev. Lett.},
  volume = {51},
  issue = {15},
  pages = {1362--1365},
  numpages = {0},
  year = {1983},
  month = {Oct},
  publisher = {American Physical Society},
  doi = {10.1103/PhysRevLett.51.1362},
  url = {https://link.aps.org/doi/10.1103/PhysRevLett.51.1362}
}

@article{Bevan1997,
    title = {Vortex Mutual Friction in Superfluid $^3${He}},
    author = {Bevan, T. D. C. and Manninen, A. J. and Cook, J. B. and Alles, H. and Hook, J. R. and Hall, H. E.},
    journal = {J. Low Temp. Phys.},
    volume = {109},
    pages = {423-459},
    year = {1997},
    doi = {10.1007/s10909-005-0095-z},
    url = {https://link.springer.com/article/10.1007/s10909-005-0095-z}
}

@phdthesis{wimanthesis,
    author = {Joshua J. Wiman},
    title = {Quantitative superfluid helium-3 from confinement to bulk},
    school = {Northwestern University},
    year = {2018}
}

@article{Barton1974,
doi = {10.1088/0022-3719/7/23/009},
url = {https://dx.doi.org/10.1088/0022-3719/7/23/009},
year = {1974},
month = {dec},
publisher = {},
volume = {7},
number = {23},
pages = {4220},
author = {G Barton and  M A Moore},
title = {Some p-wave phases of superfluid helium-3 in strong-coupling theory},
journal = {J. Phys. C: Solid State Phys.}
}

@article{Nagamura2018,
  title = {Stability of half-quantum vortices in equal-spin pairing states of $^{3}\mathrm{He}$},
  author = {Nagamura, Natsuo and Ikeda, Ryusuke},
  journal = {Phys. Rev. B},
  volume = {98},
  issue = {9},
  pages = {094524},
  numpages = {15},
  year = {2018},
  month = {Sep},
  publisher = {American Physical Society},
  doi = {10.1103/PhysRevB.98.094524},
  url = {https://link.aps.org/doi/10.1103/PhysRevB.98.094524}
}

@misc{Supplemental,
  note = "See Supplemental Material at [URL will be inserted by publisher] for computational details, explanation of various superfluid phases, discussion on the effects of Fermi-liquid corrections and the soft cores, and derivation of spin currents."
}

@article{Dalfovo1999,
  title = {Theory of {B}ose-{E}instein condensation in trapped gases},
  author = {Dalfovo, Franco and Giorgini, Stefano and Pitaevskii, Lev P. and Stringari, Sandro},
  journal = {Rev. Mod. Phys.},
  volume = {71},
  issue = {3},
  pages = {463--512},
  numpages = {0},
  year = {1999},
  month = {Apr},
  publisher = {American Physical Society},
  doi = {10.1103/RevModPhys.71.463},
  url = {https://link.aps.org/doi/10.1103/RevModPhys.71.463}
}

@article{Vorontsov2007,
  title = {Crystalline Order in Superfluid $^{3}\mathrm{He}$ Films},
  author = {Vorontsov, A. B. and Sauls, J. A.},
  journal = {Phys. Rev. Lett.},
  volume = {98},
  issue = {4},
  pages = {045301},
  numpages = {4},
  year = {2007},
  month = {Jan},
  publisher = {American Physical Society},
  doi = {10.1103/PhysRevLett.98.045301},
  url = {https://link.aps.org/doi/10.1103/PhysRevLett.98.045301}
}

@article{Wiman2016,
   doi       = {10.1007/s10909-016-1632-7},
   title     = {Strong-coupling and the stripe phase of $^3${H}e},
   author    = {Wiman, J. J. and Sauls, J. A.},
   publisher = {Springer US},
   journal   = {J. Low Temp. Phys.},
   year      = {2016},
   volume    = {184},
   pages      = {1054--1070}
}

@article{Wiman2018,
  title = {Spontaneous Helical Order of a Chiral $p$-Wave Superfluid Confined in Nanoscale Channels},
  author = {Wiman, J. J. and Sauls, J. A.},
  journal = {Phys. Rev. Lett.},
  volume = {121},
  issue = {4},
  pages = {045301},
  numpages = {6},
  year = {2018},
  month = {Jul},
  publisher = {American Physical Society},
  doi = {10.1103/PhysRevLett.121.045301},
  url = {https://link.aps.org/doi/10.1103/PhysRevLett.121.045301}
}

@article{Aoyama2014,
  title = {Stripe order in superfluid $^3${H}e confined in narrow cylinders},
  author = {Aoyama, Kazushi},
  journal = {Phys. Rev. B},
  volume = {89},
  issue = {14},
  pages = {140502(R)},
  numpages = {5},
  year = {2014},
  month = {Apr},
  publisher = {American Physical Society},
  doi = {10.1103/PhysRevB.89.140502},
  url = {https://link.aps.org/doi/10.1103/PhysRevB.89.140502}
}

\clearpage

\section*{End Matter}

\appendix

\begin{figure}
    \centering
    \includegraphics[width=\linewidth]{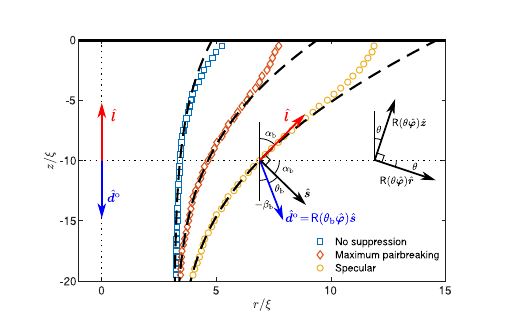}
    \caption{Comparison of simulated funnel profiles (circles) to the $r_{\rm{b}}(z)$ predicted by the model (dashed lines). The simulated data for maximum pairbreaking and specular boundaries corresponds to the states in Fig.~\ref{fig:semi-infinite}(b) and (d), respectively, and for the state with no suppression the boundary conditions were removed and the only effect arises from the abrupt end of the simulation domain. The funnel profiles are determined by calculating the order parameter distance to the bulk phase (see Appendix C of Ref.~\cite{Rantanen2024}) and taking the point where the distance crosses $0.1$. The bulk phase is defined as the order parameter state furthest from the vortex axis, separately for each $z$. The model curves are exactly the same for all three cases, shifted vertically so that the model and the data coincide at $z = -10\xi$. No fitting parameters are used otherwise.}
    \label{fig:modelcomparison}
\end{figure}

{\textit{Appendix A: Analytical model}\label{sec:amod}}---A simple picture of the A-phase-core vortex is that the vortex core consists of the A phase \eqref{eq:aphase} surrounded by the B phase \eqref{eq:bphase} outside. Here we consider a model where the interface between the two phases is abrupt. The minimal energy of the  AB interface is achieved by the condition \cite{Cross1977,Thuneberg1991}
\begin{eqnarray}
\hat{\bm d}(\hat{\bm m} +i\hat{\bm n})\cdot\hat{\bm s}=e^{i\phi}\textsf{R}\cdot\hat{\bm s},
\label{e.bocabi}\end{eqnarray}
where $\hat{\bm s}$ is the normal vector of the interface (Fig.~\ref{fig:modelcomparison}). 
This implies that $\hat{\bm l}$ has to be in the plane of the interface, $\hat{\bm l}\cdot\hat{\bm s} =0$. In order to satisfy Eq.~\eqref{e.bocabi}, we have to consider the bending of the order parameters in the two phases. 

We use cylindrical coordinates $(r,\varphi,z)$.  It is convenient to represent $\hat{\bm d}$ by an orbital-space vector  $\hat{\bm d}^{\rm o}$ so that $\hat{\bm d}=\textsf{R}(\bm\vartheta_0)\hat{\bm d}^{\rm o}$. We write the bending of the order parameters \eqref{eq:bphase} and \eqref{eq:aphase} assuming axial symmetry
\begin{eqnarray}
\bm\theta=\theta\hat{\bm\varphi},\quad 
\hat{\bm d}^{\rm o}=-p(\hat{\bm r}\sin\beta+\hat{\bm z}\cos\beta),\nonumber\\
\hat{\bm l}=\nu(\hat{\bm r}\sin\alpha+\hat{\bm z}\cos\alpha),\nonumber\\
\hat{\bm m} +i\hat{\bm n}=\textsf{R}(\alpha\hat{\bm\varphi})(\hat{\bm x} +i\nu\hat{\bm y}).
\label{e.apcpam}\end{eqnarray}
Outside the soft core $\theta\rightarrow0$. Note that $\hat{\bm l}$ is unique at the vortex axis only for $\sin\alpha=0$. 
The sign of $\hat{\bm{l}} = \nu\hat{\bm{z}}$ on the axis is correlated with the vortex circulation $\phi = \nu\varphi$ due to Eq.~\eqref{e.bocabi}. Similarly the sign of $\hat{\bm{d}}^{\rm o}$ at the axis with $\beta=0$ is determined by the parity $p$ of the vortex.
In a bulk vortex the angles $\theta$, $\beta$, and $\alpha$ are functions of $r$, but the structure close to a surface can be included allowing them to depend also on $z$. 

The requirement $\hat{\bm l}\cdot\hat{\bm s} =0$ fixes the interface normal $\hat{\bm s}=\hat{\bm r}\cos\alpha_{\rm b}-\hat{\bm z}\sin\alpha_{\rm b}$, where the subindex $\rm b$ refers to the location at the interface. 
Additionally, we have the condition $\hat{\bm d}^{\rm o}_{\rm b}=\textsf{R}(\theta_{\rm b}\hat{\bm\varphi})\cdot\hat{\bm s}$ or, equivalently,
\begin{eqnarray}
\theta_{\rm b}-\beta_{\rm b}=p\frac{\pi}2-\alpha_{\rm b},
\label{e.acbta}\end{eqnarray}
which is illustrated in Fig.~\ref{fig:modelcomparison}. In a bulk vortex $\alpha_{\rm b}=0$ due to translational symmetry along $z$ and we want to minimize the bending, that is $|\theta_{\rm b}|$ and $|\beta_{\rm b}|$.  We see that for minimal energy $\theta_{\rm b}>0$ and $\beta_{\rm b}<0$ for $p=+1$ and they have opposite signs for  $p=-1$. 
The two forms are obtained from each other by inversion, or more precisely, by reflection in the $x$-$y$ plane combined with the corresponding reflection in spin space \cite{Salomaa1985}.

Consider now the vortex ending on a surface, where $\alpha_{\rm b}$ can be non-zero. We see from Eq.~\eqref{e.acbta} that a positive $\alpha_{\rm b}$ reduces the bending proportional to  $\theta_{\rm b}$ and $-\beta_{\rm b}$ in case of $p=+1$.  In the extreme case $\alpha_{\rm b}=\pi/2$, there would be no bending of $\theta$ and $\beta$. Thus such a $p=+1$ vortex will expand at the top surface and contract at the bottom surface. The case is the opposite for $p=-1$.

We now make the model quantitative by evaluating the vortex energy on the basis of the Ginzburg-Landau gradient energy \eqref{eq:fgrad}. In case of $\alpha_{\rm b}=0$ we get the energy per vortex length $2\pi U$ with
\begin{eqnarray}
U(r_{\rm b},\theta_{\rm b},\beta_{\rm b})
=3K\Delta_{\rm B}^2\theta_{\rm b}^2+4K\Delta_{\rm A}^2\beta_{\rm b}^2\nonumber\\+5K\Delta_{\rm B}^2\ln\frac{R}{r_{\rm b}}+\sigma r_{\rm b}+{\textstyle\frac12}(f_{\rm c}^{\rm B}-f_{\rm c}^{\rm A})r_{\rm b}^2.
\label{e.poeapcv}\end{eqnarray}
Here the first and second terms are the bending energies of $\theta(r)$ and $\beta(r)$, respectively. They are evaluated to leading order in the bending angles. The third term is the kinetic energy of superflow around the vortex, where $R$ is a cut-off radius. The fourth term is the surface energy of the AB interface, where $\sigma\approx 0.9f_{\rm c}\xi$ is the surface tension \cite{Schopohl1987,Thuneberg1991}. The last term comes from the difference of the condensation energies of the A and B phases, $f_{\rm c}^{\rm A}$ and $f_{\rm c}^{\rm B}$, respectively. We use $f_{\rm c}^{\rm A}/f_{\rm c}^{\rm B}=0.95$ in the estimates below.

Minimizing Eq.~\eqref{e.poeapcv} with respect to $\theta_{\rm b}$ and $\beta_{\rm b}$ taking into account Eq.~\eqref{e.acbta} gives
\begin{eqnarray}
\theta_{\rm b}=\frac{c(p\frac{\pi}2-\alpha_{\rm b})}{1+c},\quad \beta_{\rm b}=-\frac{p\frac{\pi}2-\alpha_{\rm b}}{1+c},
\label{e.bemina}\end{eqnarray}
where $c=2f_{\rm c}^{\rm A}/f_{\rm c}^{\rm B}$. We get  $\theta_{\rm b}\approx p\frac{\pi}3$ and $\beta_{\rm b}\approx -p\frac{\pi}6$ at $\alpha_{\rm b}=0$. Further minimization with respect to $r_{\rm b}$ gives the equilibrium radius $r_{\rm{b}0}=3.1\xi$, which agrees with the numerical solution of the GL equations \cite{Thuneberg1987}.

In order to study the flaring out of the interface, we form an energy functional to determine $r_{\rm b}(z)$. For $\hat{\bm l}$ to stay in the plane of the interface  we have $r'_{\rm b}\equiv dr_{\rm b}/dz=\tan\alpha_{\rm b}\approx \alpha_{\rm b}$. We substitute $\alpha_{\rm b}=r_{\rm b}'$ to Eq.~\eqref{e.bemina}. To the energies in Eq.~\eqref{e.poeapcv} we need to add the bending energy of $\hat{\bm l}$, $U_\alpha=2K\Delta_{\rm A}^2\alpha_{\rm b}^2=2K\Delta_{\rm A}^2r_{\rm b}'^2$ in the leading order.  The AB interface energy has to be generalized to
\begin{eqnarray}
\sigma r_{\rm b}\sqrt{1+\tan^2\alpha_{\rm b}}\approx \sigma r_{\rm b}(1+{\textstyle\frac12}r_{\rm b}'^2).
\end{eqnarray} 
We neglect gradients in the $z$ direction.
For total vortex energy, the contributions above have to be integrated over $z$, 
\begin{equation}
\frac{F}{2\pi}
= \int dz\left[-\frac{3\pi pc}{1+c}K\Delta_{\rm B}^2r_{\rm b}'+M(r_{\rm b})r_{\rm b}'^2+\Delta U\right],
\label{e.fiszb}\end{equation}
where $\Delta U=U(r_{\rm b})-U(r_{\rm{b}0})$,
\begin{eqnarray}
M(r_{\rm b})=\frac{3c}{1+c}K\Delta_{\rm B}^2 +2K\Delta_{\rm A}^2+\frac12\sigma r_{\rm b},\nonumber\\
U(r_{\rm b})=5K\Delta_{\rm B}^2\ln\frac{R}{r_{\rm b}}+\sigma r_{\rm b}+\frac12(f_{\rm c}^{\rm B}-f_{\rm c}^{\rm A})r_{\rm b}^2.
\label{e.fiszinmu}\end{eqnarray}
Because of the $r_{\rm b}'$ linear term in Eq.~\eqref{e.fiszb}, the bulk vortex structure is unstable near walls. Minimization of Eq.~\eqref{e.fiszb} gives an equation $M(r_{\rm b})r_{\rm b}'^2-U(r_{\rm b})={\rm constant}$, from which $r_{\rm b}(z)$ is solved numerically. The result is in good agreement with the Ginzburg-Landau simulations, see Fig.~\ref{fig:modelcomparison}.

{\textit{Appendix B: Surface gradient contribution}}---On the surfaces of $^3$He-B there is spontaneous spin current  \cite{Zhang1987,Thuneberg2001,Tsutsumi2011,Tsutsumi2012,Wu2013,Mizushima2015,Mizushima2018}.
This wall-spin supercurrent can be written 
\begin{eqnarray}
J^{\rm ws}_{\mu i}=\frac{2m}\hbar\lambda \epsilon_{ijk}R_{\mu j}(\bm\vartheta)\hat w_k.
\end{eqnarray}
In the GL region $\lambda\approx 2.3K\Delta_B^2$ for a specular surface and  $\lambda\approx 1K\Delta_B^2$ for a diffuse surface \cite{Thuneberg2001}. (We express the spin current in mass current units. The intrinsic angular-momentum current is obtained by multiplying by the spin-mass ratio $\hbar/2m$.) The surface  flow interacts with other spin flows characterized by spin velocity
\begin{equation}
    v^{\rm s}_{\alpha i} =- \frac{\hbar}{4m}\varepsilon_{\alpha\mu\nu}R_{\mu j}\partial_i R_{\nu j}
\end{equation}
In a vortex the spatial dependence of the rotation matrix results in vortex-spin velocity
\begin{eqnarray}
\mathsf{v}^{\rm vs}(r)=\frac{\hbar}{2m}\mathsf{R}(\bm \vartheta_0)\left(\begin{array}{ccc}0& \frac1r\sin\theta & 0 \\ -\partial_r\theta& 0 & 0 \\0 & -\frac1r(1-\cos\theta)& 0\end{array}\right)
\label{e.spinvva}\end{eqnarray}
where the matrix is expressed in cylindrical coordinates $(r,\varphi, z)$. The leading contribution in the vortex spin current $\propto r^{-2}$ is of dipole form (in 2 dimensions), compared to monopole form $\propto r^{-1}$ of the mass current. The spin current is conserved as long as we can neglect the spin-orbit interaction, $r\ll 10\,\text{\textmu m}$.

The integrated interaction energy with the wall-spin current is
\begin{eqnarray}
\frac{F_{\rm spin}}{2\pi}
=\int_{r_b}^{\infty}dr\, r\,v^{\rm vs}_{\alpha i}J^{\rm ws}_{\alpha i}
\approx\frac1{12}\lambda\theta_b^3 r_b\hat w_z.
\label{e.funnelsuf2}
\end{eqnarray}
Depending on the $\operatorname{sign} \theta_b^3 \hat w_z  = \operatorname{sign} \hat w_z p$ (that is, at the left- or right-handed vortex end), this energy is reduced by the increase or decrease of the core radius $r_{\rm b}$. This contribution is in the same direction as the effect of flaring out the interface in Appendix A. Minimization of the total energy from Eqs.~\eqref{e.poeapcv} and \eqref{e.funnelsuf2} with respect to $r_{\rm{b}}$ for $p=+1$ gives $r_{\rm{b}}\approx 3.5\xi$ at the left-handed end and $r_{\rm{b}}\approx 2.8\xi$ at the right-handed end. Note that the surface term is only effective in the suppression layer extending a few coherence lengths from the boundary.

{\textit{Appendix C: Implications for the vortex-core transition in bulk $^3$He}}---Many features of the transition between A-phase-core and double-core vortices observed in experiments \cite{Hakonen1983,Pekola1984,Kondo1991,Bevan1997} are still lacking encompassing explanation. Based on calculations in the GL region \cite{Thuneberg1987} it was interpreted as a thermodynamic transition, while recently it was suggested that the transition rather represents the end of the metastability region of the A-phase-core vortex~\cite{Regan2020,Rantanen2024}. 

Detailed implications of our findings for those interpretations should be considered elsewhere. Here we list a few observations:
First, both core types being present at different ends of a vortex (Fig.~\ref{fig:semi-infinite}b and c) implies that the possible nucleation barrier masking the thermodynamic transition is small. Second, the A-phase funnel affects the energy balance of the vortices and thus could provide a reason for the difference in the vortex core transition observed with different vortex lengths \cite{Pekola1984,Bevan1997}. Third, in the case that the double-funnel structure (Fig.~\ref{fig:semi-infinite}d) would be present, it would increase the metastability of the A-phase-core vortex.

\newpage

\setcounter{equation}{0}
\renewcommand{\theequation}{S\arabic{equation}}

\title{Supplemental material \\Surface-induced vortex core restructuring in a spin-triplet superfluid}%
\maketitle
\onecolumngrid

\section{Computational details}
All simulation results in the main text are obtained by numerically minimizing the Ginzburg-Landau free energy in the system with appropriate boundary conditions and initial states. The domain is discretized into a mesh of tetrahedral elements, with the full complex $3\times 3$ order parameter matrix defined at the vertices of the mesh. The free energy is calculated in this discretized basis~(see Ref.~\cite{Rantanen2025a} for details) and minimized using the Limited-memory Broyden-Fletcher-Goldfarb-Shanno (L-BFGS) algorithm.

The simulations are performed in a three-dimensional cylindrical box, with slightly different sizes and resolutions for different results. The structures in Fig.~1 of the main text are calculated in a cylinder with radius of $35\xi$, height $75\xi$ and a resolution of $0.5\xi$ in all directions. The vortices in Fig.~2 of the main text use cylinders with a larger radius of $400\xi$ in order to account for the soft cores. The heights of the two slabs are 700~nm and 1500~nm, and $70\xi$ for the semi-infinite case. The resolution in the $xy$ plane is $0.25\xi$ up to a $15\xi$ radius, beyond which it gradually increases to about $20\xi$ at $400\xi$. The resolution in the $z$ direction is 10~nm, 20~nm, and $\xi$ for the 700~nm, 1500~nm, and semi-infinite case, respectively. The hysteresis curve of Fig.~3 in the main text is calculated in a 700~nm high cylinder with a radius of 717~nm, and a resolution of 10~nm in all directions.

The boundary condition on the top and bottom of the cylinder depend on specularity (explained below) for solid surfaces. For the semi-infinite calculations, on the side of the cylinder representing the bulk the order parameter is fixed to the vortex structure obtained from bulk calculations~\cite{Rantanen2024}.

In the absence of a stabilizing force such as external rotation, the vortex is pulled towards the sides of the calculation box and eventually leaves the system. In order to prevent this we apply a boundary condition on the radial boundary of the cylinder that enforces a 180$^\circ$ rotational symmetry. This is done by setting the order parameter at each point $\bm{r}_1$ on the boundary to be equal to the order parameter on the opposite side $\bm{r}_2$ of the cylinder, rotated and phase shifted by $\pi$:
\begin{equation}
    \textsf{A}(\bm{r}_1) = e^{i\pi}\textsf{R}(\pi\hat{\bm{z}})\textsf{A}(\bm{r}_2)\textsf{R}^T(\pi\hat{\bm{z}}).
\end{equation}
The same condition was used previously to calculate the structure of vortices in the bulk~\cite{Rantanen2024}, as it doesn't require \textit{a priori} knowledge of the soft core structure and minimally constrains the possible core structures. Note that in the calculations done with a larger radius of $400\xi$, this boundary condition is not applied as the vortex remains in the middle of the computation box without it.

We use two common boundary conditions for the surface: maximum pairbreaking with complete suppression of the order parameter and fully specular conditions with suppression of only the transverse component~\cite{Wiman2015}. Maximum pairbreaking is enforced by setting the order parameter at the surface to zero, $\textsf{A} = 0$. Specular boundary conditions set the perpendicular orbital components to zero, $A_{\mu z} = 0$, and also prevent gradients through the boundary, $\nabla_z A_{\mu x} = \nabla_z A_{\mu y} = 0$. The specular conditions are enforced through a penalty term in the energy.

The temperature and pressure dependence of our calculations comes from the Ginzburg-Landau coefficients. In addition to the usual GL parameters, we use temperature dependent strong-coupling corrections for the quartic $\beta$ coefficients~\cite{Regan2020}. The corrections are implemented as
\begin{equation}
    \beta_i(p,T) = \beta_i^{\mathrm{wc}}(p) + \frac{T}{T_{\mathrm{c}}}\beta_i^{\mathrm{sc}}(p)
    \label{supp:sc}
\end{equation}
where $\beta_i^{\mathrm{wc}}$ are the weak-coupling values and $\beta_i^{\mathrm{sc}}$ are the strong-coupling corrections from Ref.~\cite{wimanthesis}, tabulated in Ref.~\cite{Rantanen2024}.

\section{Superfluid phases of $^3$He}
The superfluid phases of $^3$He are the uniform order parameter solutions of the Ginzburg-Landau equations defined by the bulk free-energy density
\begin{align}
    f_{\text{bulk}} &= \alpha\Tr\!\left\{AA^\dag\right\} + \beta_1|\Tr\!\left\{AA^T\right\}|^2 + \beta_2\Tr\!\left\{AA^\dag\right\}^2 \label{supp:fbulk} \\ &+ \beta_3\Tr\!\left\{AA^T(AA^T)^*\right\} + \beta_4\Tr\!\left\{(AA^\dag)^2\right\} + \beta_5\Tr\!\left\{AA^\dag(AA^\dag)^*\right\} \nonumber
\end{align}
The lowest energy phase is determined by the values of the $\beta_i$ parameters. For an order parameter with amplitude $\Delta$, the fourth-order terms in Eq.~\eqref{supp:fbulk} can be written as $\Delta^4\sum_{i=1}^5 \beta_i I_i$, with $I_i$ having values between 0 and 1. 

According to the classification by Barton and Moore~\cite{Barton1974}, there are seven so-called inert phases of superfluid $^3$He, i.e. solutions of the GL equations that are independent of the $\beta_i$ parameters up to overall amplitude $\Delta$. When $\Delta$ is excluded by introducing quantities $I_i$ as shown above, each inert phase is represented by a unique set of fixed values of $I_i$, now called fourth-order invariants, which can be used as an identifier for that phase. The values of $I_i$ for each phase is listed in Table~\ref{supp:invariants}.

\begin{table}[h]
\centering
\caption{\label{supp:invariants}%
The values of the fourth-order invariants for each inert phase. Note that the A$_1$ phase was labeled $\gamma$ in Ref.~\cite{Barton1974}.}
\begin{tabular}{c c c c c c}
\hline \hline
\textrm{Phase}&
$\quad I_1\quad$&
$\quad I_2\quad$&
$\quad I_3\quad$&
$\quad I_4\quad$&
$\quad I_5\quad$\\ \hline
\textrm{B} & 1 & 1 & 1/3 & 1/3 & 1/3 \\
\textrm{Planar} & 1 & 1 & 1/2 & 1/2 & 1/2 \\
\textrm{Polar} & 1 & 1 & 1 & 1 & 1 \\
$\alpha$ & 0 & 1 & 1/3 & 1/3 & 1/3 \\
\textrm{Bipolar} & 0 & 1 & 1/2 & 1/2 & 1/2 \\
\textrm{A} & 0 & 1 & 0 & 1 & 1 \\
$\beta$ & 0 & 1 & 1 & 1 & 0 \\
A$_1 (\gamma)$ & 0 & 1 & 0 & 1 & 0 \\ 
\hline \hline
\end{tabular}
\end{table}

Out of the phases listed in Table~\ref{supp:invariants}, the A, B, and A$_1$ phases are found in bulk superfluid $^3$He. The polar and $\beta$ phases have been observed in anisotropic confinement. While some phases have not been observed in a uniform system, they can appear inside vortex cores. We list the order parameters of the phases mentioned in the main text below.

The planar phase is not stable in the bulk, but is found at the boundaries of the B phase where the gap is suppressed in one direction. The order parameter for the planar phase is
\begin{equation}
    A_{\mu j} = \Delta_{\rm P} e^{i\phi}R_{\mu k}(\delta_{kj} - \hat{p}_k\hat{p}_j)
    \label{supp:planar}
\end{equation}
where $\hat{\bm{p}}$ indicates the direction of the suppression.

The $\beta$ phase is the spin-polarized version of the polar phase, like the A$_1$ phase is to the A phase. Its order parameter is
\begin{equation}
    A_{\mu j} = \Delta_\beta (\hat{d}_\mu + i\hat{e}_\mu)\hat{m}_j
    \label{supp:beta}
\end{equation}
where $\hat{\bm{d}}$ and $\hat{\bm{e}}$ are orthogonal vectors in spin space and $\hat{\bm{m}}$ is a vector in orbital space.

The $\alpha$ phase has not been observed in the bulk, but simulations of the double-core vortex show that a state resembling it appears inside the two subcores. An example of the $\alpha$ phase order parameter is
\begin{equation}
    \textsf{A} = e^{i\phi}\Delta_\alpha \begin{bmatrix}
        1 & 0 & 0 \\
        0 & e^{i\pi/3} & 0 \\
        0 & 0 & e^{i2\pi/3}
    \end{bmatrix}
    \label{supp:alpha}
\end{equation}

The axiplanar state is not listed in Table~\ref{supp:invariants}, as it is non-inert. However, it is found at interfaces between the A and B phases, such as the one surrounding the A-phase-core vortex. The axiplanar order parameter is
\begin{equation}
    A_{\mu j} = \frac{1}{2}\Delta_\uparrow (\hat{\bm{d}} + i\hat{\bm{e}})_\mu(\hat{\bm{m}}_\uparrow + i\hat{\bm{n}}_\uparrow)_j + \frac{1}{2}\Delta_\downarrow (\hat{\bm{d}} - i\hat{\bm{e}})_\mu(\hat{\bm{m}}_\downarrow + i\hat{\bm{n}}_\downarrow)_j
\end{equation}
where $\uparrow$ and $\downarrow$ indicate spin-up and spin-down components, respectively. The two spin components have different orbital orientations, $\hat{\bm{l}}_\uparrow = \hat{\bm{m}}_\uparrow \times \hat{\bm{n}}_\uparrow$ and $\hat{\bm{l}}_\downarrow = \hat{\bm{m}}_\downarrow \times \hat{\bm{n}}_\downarrow$, and the axiplanar state can be seen as an interpolation between the A phase with $\hat{\bm{l}}_\uparrow = \hat{\bm{l}}_\downarrow$ and the planar phase with $\hat{\bm{l}}_\uparrow = -\hat{\bm{l}}_\downarrow$.

Using the invariant signatures for each phase listed in Table~\ref{supp:invariants}, we determine the nearest inert phase by calculating the fourth-order invariants for a given order parameter $\textsf{A}$ and comparing the distance to each phase. For example, the distance to the B phase is given by
\begin{equation}
    D_{\rm B} = \sqrt{\sum_{i=1}^5\left(I_i[\textsf{A}]-I_i^{\rm B}\right)^2}
\end{equation}
where $I_i^{\rm{B}}$ are the invariant values of the B phase listed in Table~\ref{supp:invariants}. The nearest phase is determined as the phase with smallest distance $D$. In regions where suppression is strong, such as near the maximum pairbreaking surfaces in Fig.~1 of the main text, the phase determination is less accurate due to the small amplitude of $\textsf{A}$.

The order parameter inside vortex cores can often be in non-inert states as well. For example, the order parameter between the planar and A phases in the A-phase-core vortex is axiplanar, while the nearest inert phase is the bipolar phase. In the main text we have labeled these regions as axiplanar, to more accurately describe the order parameter structure.

\section{Fermi-liquid corrections}

The Fermi-liquid interaction parameters are included in the coefficients of the standard Ginzburg-Landau theory of superfluid $^3$He. We use an extended GL theory that adds temperature dependence in the quartic coefficients (see Eq.~\eqref{supp:sc}) as the only extension compared to the standard theory. This captures some effects of Fermi-liquid interactions, but misses other changes in the energy functional that formally affect the temperature dependence in the same order. A fuller account for the Fermi-liquid effects in the GL model is possible by including higher-order gradient terms in the energy functional~\cite{Nagamura2018}.

In $^3$He the Fermi-liquid interactions are especially strong in the density response and mass-current response, parametrized by $F_0^s$ and $F_1^s$, respectively. In the stationary case $F_0^s$ does not appear, but the large value of $F_1^s$ is important. With respect to vortex structures, a major effect from Fermi-liquid corrections is that at lower temperatures gradients that cause mass current become much more energetically expensive than other gradients. That changes the balance between the mass and spin supercurrents, which in the GL theory carry the same superfluid density. A known consequence of this is an increase of the distance between subcores of the double-core vortex by a factor of 2--3 compared to pure GL calculations~\cite{Silaev15}.

Based on this, we expect that the true funnel radius could be larger than our model suggests because the mass flow is reduced in the expanded vortex end. The opposite effect appears in the contracting end because it leads to increased mass flow. 
Because other Fermi-liquid interaction parameters, as far as they are known, are small compared to $F_1^s$, we do not expect essential changes in properties that do not cause or affect mass flow.

Thus, we expect that taking Fermi-liquid corrections into account will change quantitative characteristics of the calculated structures, e.g. the size of the funnel. However, we believe that qualitatively the observed structures are robust, since they do not depend on fine-tuning of the interaction parameters. This is confirmed in particular by consistent observation of the A-phase funnel at all pressures and temperatures. Additionally, at the low temperatures where Fermi-liquid corrections are the most important, GL theory is expected to provide only qualitative results in any case. 

\section{Effect of the soft core}

\begin{figure}
    \centering
    \includegraphics[width=0.8\linewidth]{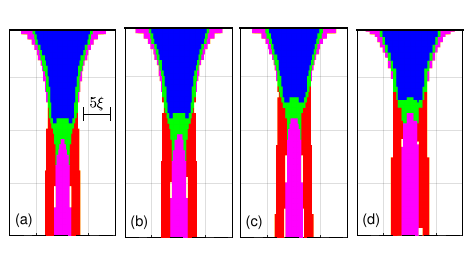}
    \caption{The calculated vortex structures at a maximum pairbreaking surface for different soft cores, with $p=20\text{ bar}$ and $T = 0.80T_{\rm{c}}$. (a) Standard spin-orbit coupling. The state corresponds to the structure in Fig~2a of the main text. (b) No spin-orbit coupling. (c) 1600 times stronger than standard spin-orbit coupling. Cases (a)-(c) are calculated in a box with $400\xi$ radius. (d) Standard spin-orbit coupling, but calculated in a box with $35\xi$ radius, which doesn't fully account for the soft-core structure.}
    \label{fig:suppsoftcores}
\end{figure}
As discussed in the main text, the surface-induced core restructuring is determined by the parity of the vortex, which also defines the orientation of the soft core. In the soft-core region the rotation matrix $\textsf{R}$ of the B phase slowly recovers the bulk value $\textsf{R}(\bm{\vartheta}_0)$ on a length scale of the dipole length $\xi_{\rm{d}} = \sqrt{K/g_{\rm{d}}}$, determined by the gradient energy coefficient $K$ and the strength of the spin-orbit interaction $g_{\rm{d}}$. While the spin-orbit interaction is important for the long-range decay of the soft core, it does not play a role in determining the parity of the vortex. In order to confirm this, we calculate the funnel structure with $g_{\rm{d}}$ set to zero, shown in Supplemental Figure~\ref{fig:suppsoftcores}b.

In solid superconducting systems the spin-orbit coupling is often much stronger than in superfluid $^3$He. In order to test the effects of stronger spin-orbit coupling, we also calculate the vortex structure with a dipole energy coefficient 1600 times its natural value, shown in Supplemental Figure~\ref{fig:suppsoftcores}c. This decreases the dipole length by a factor of 40, making it comparable to the size of the vortex core. The qualitative surface structure remains the same, but the size of the A-phase-core segment is slightly reduced.

A smaller calculation domain does not take into account the slow decay of $\theta$ in the soft core, which also results in a reduced size for the A-phase-core segment, shown in Supplemental Figure~\ref{fig:suppsoftcores}d.

\section{Spin currents in the B phase}
The flow of spin in $^3$He-B is described by the spin current density
\begin{equation}
    j^{\rm{sp}}_{\mu i} = \frac{\partial f}{\partial \omega_{\mu i}}
    \label{supp:js1}
\end{equation}
where $f$ is the free-energy density and $\omega_{\mu i} = \frac{1}{2}\varepsilon_{\mu\nu\lambda}R_{\nu j}(\partial_i R_{\lambda j})$ is the spin connection, commonly referred to as the spin superfluid velocity in $^3$He literature (up to the sign and coefficient, depending on desired units). In order to calculate $\bm{j}^{\rm{sp}}_\mu$ in the B phase, we start from the general (distorted) B phase order parameter
\begin{equation}
    \mathsf{A} = \mathsf{R}\begin{bmatrix}
        \Delta_x & 0 & 0 \\
        0 & \Delta_y & 0 \\
        0 & 0 & \Delta_z
    \end{bmatrix}
    \label{supp:Bop}
\end{equation}
where $\Delta_i$ is the size of the gap in the $i$ direction ($\Delta_x = \Delta_y = \Delta_z$ in the bulk). Next we can express the gradient energy density in terms of $\omega_{\mu i}$ by inserting Eq.~\eqref{supp:Bop} into Eq.~(3) in the main text:
\begin{align}
    f_{\text{grad}} &= K_1\left[ 2 \omega_{\lambda i}\omega_{\lambda i} \Delta_j^2 + 3\left(\nabla_i \Delta_j\right)^2 \right] \nonumber \\ &+ (K_2 + K_3)\left[ \omega_{\lambda i}\omega_{\lambda i}\Delta_i^2 - \omega_{\lambda i}\omega_{\nu j} R_{\nu i}R_{\lambda j} \Delta_i \Delta_j + \left(\nabla_i\Delta_i\right)^2 \right] \nonumber \\
    &+ 2\varepsilon_{ijk}\omega_{\lambda i}R_{\lambda k}\left[K_2\Delta_i\left(\nabla_j\Delta_j\right) - K_3\Delta_j\left(\nabla_j\Delta_i\right)\right]
    \label{supp:fgrad}
\end{align}
where we made use of the expression $\partial_i R_{\mu k} = -\varepsilon_{\mu\nu\lambda}\omega_{\lambda i}R_{\nu k}$. From this we get the spin current density for the B phase
\begin{align}
    j^{\rm{sp}}_{\mu i} &= 4K_1\Delta_j^2 \omega_{\mu i} + (K_2 + K_3)\left[ 2\Delta_i^2 \omega_{\mu i} - 2\Delta_i\Delta_j\omega_{\nu j} R_{\nu i}R_{\mu j} \right] \nonumber \\
    &+ 2\varepsilon_{ijk} R_{\mu k}\left[K_2\Delta_i\left(\nabla_j \Delta_j\right) - K_3\Delta_j\left(\nabla_j\Delta_i\right) \right].
    \label{supp:js2}
\end{align}
From Eq.~\eqref{supp:js2} we see that $j_{\mu i}^{\rm{sp}}$ contains two contributions: the current $j_{\mu i}^{\text{sp,text}}$ from the spatial variation (texture) of $\textsf{R}$ (first line of Eq.~\eqref{supp:js2}), and the spontaneous spin currents $j_{\mu i}^{\text{sp,spont}}$ from suppression of $\Delta$ (second line of Eq.~\eqref{supp:js2}). When integrated over the suppression region near the surface, the latter produces the well-known surface spin currents in the B phase, Eq.~(12) in the main text. The dependence of $\lambda$ in that equation on the boundary conditions comes from the gradients of $\Delta$ in $j_{\mu i}^{\text{sp,spont}}$ being determined by those conditions. The interaction energy of the surface spontaneous spin current and the spin connection is in the last line of Eq.~\eqref{supp:fgrad}, $f_{\text{spin}} = j_{\mu i}^{\text{sp,spont}}\omega_{\mu i}$. After integration it leads to Eq.~(15) in the main text.

\end{document}